# Emissive Azobenzenes Delivered on a Silver Coordination Polymer

*Jingjing Yan,[†] Liam Wilbraham,[‡,§] Prem N. Basa,[†] Mischa Shuettel,[†] John C. MacDonald,[†] Ilaria Ciofini,[‡] François-Xavier Coudert[‡] and Shawn C. Burdette*,[†]*

[†]Department of Chemistry and Biochemistry, Worcester Polytechnic Institute, 100 Institute Road, Worcester, MA 01609-2280 (USA) and [‡]Chimie ParisTech, PSL Research University, CNRS, Institut de Recherche de Chimie Paris, 75005 Paris, France

**ABSTRACT**

Azobenzene has become a ubiquitous component of functional molecules and polymeric materials because of the light-induced *trans*→*cis* isomerization of the diazene group. In contrast, there are very few applications utilizing azobenzene luminescence since excitation energy typically dissipates via non-radiative pathways. Inspired by our earlier studies with 2,2′-bis[N,N′-(2-pyridyl)methyl]diaminoazobenzene (AzoAM*o*P) and related compounds, we investigated a series of five aminoazobenzene derivatives and their corresponding silver complexes. Four of the aminoazobenzene ligands, which exhibit no emission under ambient conditions, form silver coordination polymers that are luminescent at room temperature. AzoAE*p*P (2,2′-bis[N,N′-(4-pyridyl)ethyl]diaminoazobenzene) assembles into a three-dimensional coordination polymer (AgAAE*p*P) that undergoes a reversible loss of emission upon the addition of metal coordinating analytes like pyridine. The switching behavior is consistent with the disassembly-reassembly of the coordination polymer driven by displacement of the aminoazobenzene ligands by coordinating analytes.

## INTRODUCTION

Azobenzene (AB) undergoes *trans*→*cis* isomerization when irradiated, and *cis*→*trans* isomerization thermally or upon exposure to a different wavelength of light. Since light absorption depends on the conjugated π-orbital system, ring substituents change the photoisomerization process and the optical properties of the AB chromophore. The $S_2 \leftarrow S_0$ of aminoazobenzene (aAB) shifts to longer wavelengths, which typically leads to an overlap with the $S_1 \leftarrow S_0$ transition.[1-3] As we previously reported, AzoAM*o*P (**1**, 2,2′-bis[*N,N′*-(2-pyridyl)methyl]diaminoazobenzene) exhibits overlapping $S_2 \leftarrow S_0$ and $S_1 \leftarrow S_0$ transitions with a $\lambda_{max}$ at 490 nm and 30-fold higher emission than AB at 77 K. Furthermore, AzoAM*o*P undergoes minimal *trans*→*cis* isomerization due to intramolecular hydrogen bonding between the anilino protons and the pyridyl and diazene nitrogen atoms.[4] Hydrogen bonding imposes an energetic barrier that prevents the aryl rings from adopting the prerequisite collinear conformation necessary for isomerization via the concerted inversion mechanism.

To further develop the photochemistry of these unique aAB derivatives, we replaced the pyridine ligand of the aminomethylpyridine groups with a series of hydrogen bond acceptors, and changed the linker between the amine and pyridine to an ethylene.[5] Investigations with the small library of compounds suggested that reproducing the structure-induced photophysical properties observed with AzoAM*o*P would be difficult with a standalone AB chromophore. This study included the report of AzoAE*o*P (**2**, 2,2′-bis[*N*,*N′*-(2-pyridyl)ethyl]diaminoazobenzene), which utilizes a pyridine ligand and an ethylene spacer. With AzoAM*o*P and AzoAE*o*P in hand, we reasoned that the using a 3-pyridyl (*meta*) or 4-pyridyl (*para*) pyridine ligand in the existing aminoazobenzene scaffold would allow retention of the hydrogen bonded anilino-hydrogen-diazene core, while orienting the pyridine nitrogen atoms on a trajectory favorable for coordinating metal ions. Since the low temperature emission of AzoAM*o*P was attributed to hydrogen bonding and from being embedded in a frozen solvent glass, we hypothesized that integrating the aAB chromophore into coordination polymers would restrict non-radiative decay pathways sufficiently to produce emissive systems at ambient temperature.

## EXPERIMENTAL SECTION

**General Procedures.** All reagents were purchased and used without further purification. The 2,2'-diaminoazobenzene (DAAB) synthon was prepared as previously described.[6] Toluene, dichloromethane ($CH_2Cl_2$), dichloroethane (DCE) and diethylether ($Et_2O$) were sparged with argon and dried by passing through alumina-based drying columns. All chromatography and thin-layer chromatography (TLC) were performed on silica (200–400 mesh). TLCs were developed by using $CH_2Cl_2$ or solvent mixtures containing $CH_2Cl_2$, ethyl acetate (EtOAc), hexanes, or methanol ($CH_3OH$). $^1H$ and $^{13}C$ NMR spectra were recorded with a 500 MHz Bruker Biospin NMR instrument. Chemical shifts are reported in ppm relative to tetramethylsilane (TMS). FT-IR spectra were recorded using Bruker Optics FT-IR spectrometer equipped with a Vertex70 attenuated total reflection (ATR) accessory by collecting 1024 scans over a scan range from 4000 to 400 $cm^{-1}$ at 4

cm$^{-1}$ resolution. Thermogravimetric analysis (TGA) measurements were carried out on a TA Instruments Hi-Res TGA 2950 Thermogravimetric Analyzer from room temperature to 800 °C under nitrogen atmosphere at a heating rate of 10 °C/min. LC/MS was carried on a Single Quadruple, Agilent Technologies 1200 series LC system. High resolution mass spectra were obtained at the University of Notre Dame mass spectrometry facility using microTOF instrument operating in positive ionization mode. Melting-point information was obtained using a Hydrothermal Mel-Temp instrument.

**2,2′-Bis[N,N′-(2-pyridyl)methyl]diaminoazobenzene (1, AzoAM*o*P).** AzoAM*o*P was synthesized as previously reported[4] with minor modifications. DAAB (0.640 g, 3.02 mmol), 2-pyridinecarboxaldehyde (0.600 mL, 6.31 mmol) and 3 Å molecular sieves (0.940 g) were combined in $CH_2Cl_2$ (30 mL) and stirred for 24 h at room temperature. Sodium triacetoxyborohydride ($NaBH(OAc)_3$, 1.34 g, 6.32 mmol) was added, and the mixture was stirred at room temperature for 24 h. The reaction mixture was diluted with water (20 mL), and the product was extracted into $CH_2Cl_2$ (3 × 40 mL). The combined organic layers were dried over sodium sulfate ($Na_2SO_4$), and the solvent was removed. Flash chromatography on silica (24:1 EtOAc/MeOH) yielded an orange-red powder (387 mg, 32.5%). Analytical data matched previously reported values.[4]

**2,2′-Bis[N,N′-(2-pyridyl)ethyl]diaminoazobenzene (2, AzoAE*o*P).** AzoAE*o*P was synthesized as previously reported[5] with minor modifications. DAAB (0.640 g, 3.02 mmol) and 2-vinylpyridine (316 μL, 2.93 mmol) were combined in $CH_3OH$ (3 mL) and acetic acid (AcOH, 10 mL). The reaction mixture was stirred for 3 h at 45 °C before a second potion of 2-vinylpyridine (632 μL, 5.86 mmol) was added. After stirring for 24 h at 45 °C, the reaction mixture was cooled to room temperature, diluted with 10 mL of water, and the pH was adjusted to ~8 with ammonium hydroxide ($NH_4OH$). The product was extracted into EtOAc (3 × 50 mL), the combined organic materials were dried over $Na_2SO_4$, and the solvent was removed. Flash chromatography on silica ($CH_2Cl_2$/$CH_3OH$, 25:1) gave AzoAE*o*P (448 mg, 35.1%) as a dark red solid. Analytical data matched previously reported values.[5]

**2,2′-Bis[N,N′-(3-pyridyl)methyl]diaminoazobenzene (3, AzoAM*m*P).** DAAB (0.640 g, 3.02 mmol), 3-pyridinecarboxaldehyde (600 µL, 6.39 mmol) and 3 Å molecular sieves (0.940 g) were combined in DCE (30 mL) and stirred for 24 h at room temperature. $NaBH(OAc)_3$ (1.34 g, 6.32 mmol) was added and the reaction mixture was stirred at room temperature for 24 h. The reaction mixture was diluted with water (20 mL), and the product was extracted into $CH_2Cl_2$ (3 × 40 mL). The combined organic layers were dried over sodium $Na_2SO_4$, and the solvent was removed. Flash chromatography on silica using EtOAc/hexanes (2:1) and EtOAc/$CH_3OH$ (20:1) yielded AzoAM*m*P as a dark red solid (532 mg, 44.7%). Diffusion of $Et_2O$ into an $CH_3CN$ solution of AzoAM*m*P provided orange-red blocks suitable for X-ray analysis. TLC $R_f$ = 0.30 (silica, EtOAc/$CH_3OH$, 49:1). Mp = 158-159 °C. $^1$H NMR (500 MHz, $CDCl_3$) δ 8.65 (s, 2 H), 8.53 (d, *J* = 4.9 Hz, 2 H), 8.43 (s, 2 H), 7.67 (d, *J* = 7.8 Hz, 2 H), 7.55 (dd, *J* = 8.0, 1.6 Hz, 2 H), 7.25 (t, *J* = 6.2 Hz, 2 H), 7.19 (t, *J* = 7.8 Hz, 2 H), 6.76 (t, *J* = 7.6 Hz, 2 H), 6.70 (d, *J* = 8.4 Hz, 2 H), 4.50, (s, 4 H). $^{13}$C NMR (125 MHz, $CDCl_3$) δ 149.1, 149.0, 143.0, 136.6, 134.8, 134.3, 131.7, 127.3, 123.6, 116.8, 112.0, 44.7. FT-IR (neat, cm$^{-1}$) 3302.0, 3062.1, 3030.0, 2991.1, 2880.2, 2617.1, 2056.7, 1498.0, 1476.3, 1465.2, 1419.9, 1309.2, 1247.5, 1199.2, 1153.9, 1122.6, 1025.9, 1041.0, 906.9, 787.0, 748.0, 706.7, 601.8. HRMS (+ESI) calculated for $MH^+$ 395.1979 and observed 395.1980.

**2,2′-Bis[N,N′-(4-pyridyl)methyl]diaminoazobenzene) (4, AzoAM*p*P).** DAAB (0.640 g, 3.02 mmol), 4-pyridinecarboxaldehyde (600 µL, 6.37 mmol) and 3 Å molecular sieves (0.940 g) were combined in DCE (30 mL) and stirred for 24 h at room temperature. $NaBH(OAc)_3$ (1.34 g, 6.32 mmol) was added and the reaction mixture was stirred at room temperature for 24 h. The reaction mixture was diluted with water (20 mL), and the product was extracted into $CH_2Cl_2$ (3 × 40 mL). The combined organic layers were dried over $Na_2SO_4$, and the solvent was removed. Flash chromatography on silica using $CH_2Cl_2/CH_3OH$ (10:1) yielded AzoAM*p*P as a red orange solid (474 mg, 39.8%). Slow evaporation of chloroform ($CHCl_3$) provided orange-red needles suitable for X-ray analysis. TLC $R_f = 0.20$ (silica, $CH_2Cl_2/CH_3OH$, 24:1). Mp = 160-161 °C. $^1$H NMR (500 MHz, $CDCl_3$) δ 8.58 (d, *J* = 5.7 Hz, 4 H), 8.55 (s, 2 H), 7.63 (dd, *J* = 7.9, 1.6 Hz, 2 H), 7.33 (d, *J* = 6.0 Hz, 4 H), 7.19 (t, *J* = 7.7 Hz, 2 H), 6.79 (t, *J* = 7.6 Hz, 2 H), 6.61 (d, *J* = 8.5 Hz, 2 H), 4.54 (s, 4 H). $^{13}$C NMR (125 MHz, $CDCl_3$) δ 150.2, 148.2, 143.0, 136.6, 131.8, 127.3, 122.0, 116.9, 112.1, 46.1. FT-IR (neat, $cm^{-1}$) 3358.0, 3111.4, 2892.3, 2130.1, 1734.2, 1652.8, 1615.2, 1528.5, 1492.4, 1415.2, 1370.5, 1312.1, 1300.9, 1244.1, 1289.1, 1206.4, 1148.1, 1124.5, 1045.8, 1016.7, 987.5, 941.5, 883.6, 804.2, 740.7, 666.3, 620.0. HRMS (+ESI) calculated for $MH^+$ 395.1979 and observed 395.1985.

**2,2′-Bis[N,N′-(4-pyridyl)ethyl]diaminoazobenzene (5, AzoAE*p*P)**. DAAB (0.640 g, 3.02 mmol) and 4-vinylpyridine (316 μL, 2.96 mmol) were combined in $CH_3OH$ (3 mL) and acetic acid (AcOH, 10 mL). The reaction mixture was stirred for 3 h at 45 °C before a second potion of 4-vinylpyridine (632 μL, 5.91 mmol) was added. After stirring for 24 h at 45 °C, the reaction mixture was cooled to room temperature, diluted with 10 mL of water, and the pH was adjusted to ~8 with $NH_4OH$. The product was extracted into EtOAc (3 × 50 mL), the combined organic materials were dried over $Na_2SO_4$, and the solvent was removed. Flash chromatography on silica using $CH_2Cl_2/CH_3OH$ (15:1) yielded AzoAE*p*P as a red orange solid (500 mg, yield 39.1%). Slow evaporation from toluene/ethanol (1:1) provided orange-red needles suitable for X-ray analysis.

TLC $R_f$ = 0.35 (silica, DCM/$CH_3OH$, 10:1). Mp = 149–150 °C. $^1$H NMR (500 MHz, $CDCl_3$) δ 8.51 (d, *J* = 4.4 Hz, 4 H), 8.06 (s, 2 H), 7.32 (dd, *J* = 8.0, 1.6 Hz, 2 H), 7.24 (t, *J* = 7.8 Hz, 2 H), 7.17 (d, *J* = 4.7 Hz, 4 H), 6.79 (d, *J* = 8.4 Hz, 2 H), 6.74 (t, *J* = 7.5 Hz, 2 H), 3.57 (q, *J* = 6.4 Hz, 4 H), 2.97 (t, *J* = 6.9 Hz, 4 H). $^{13}$C NMR (125 MHz, $CDCl_3$) δ 150.0, 148.1, 143.0, 136.4, 131.6, 127.2, 124.1, 116.3, 111.7, 42.9, 34.8. FT-IR (neat, $cm^{-1}$) 3205.4, 3046.8, 2917.4, 2851.4, 2171.9, 2330.4, 1601.0, 1565.1, 1509.1, 1458.6, 1321.6, 1204.0, 1147.8, 1072.3, 1042.8, 836.7, 785.9, 368.8. HRMS (+ESI) calculated for $MH^+$ 423.2292 and observed 423.2270.

**{[Ag(AzoAM*o*P)]($CF_3SO_3$)($CH_3CN$)}$_n$ (AgAAM*o*P).** AzoAM*o*P (25.4 mg, 64.4 μmol) in toluene (1.8 mL) was added dropwise to a toluene solution (1.8 mL) of silver trifluoromethanesulfonate (AgOTf, 16.6 mg, 64.6 μmol). Upon stirring the reaction mixture for 30 min, an orange-red solid slowly precipitated. $CH_3CN$ (1 mL) was added to re-dissolve the precipitate, and the reaction mixture was stirred at room temperature for 2 h and filtered. Slow evaporation provided orange rectangular plates suitable for X-ray analysis. FT-IR (neat, $cm^{-1}$) 3361.6, 3069.5, 2904.1, 2324.2, 1981.4, 1597.2, 1581.2, 1493.6, 1466.0, 1436.7, 1371.0, 1322.3, 1285.2, 1240.9, 1219.3, 1159.1, 1107.3, 1075.2, 1052.2, 1028.3, 991.5, 888.3, 823.5, 755.2, 699.3, 633.5. Elemental analysis calcd. for AgAAMoP $C_{27}H_{25}AgF_3N_7O_3S$: C 46.79%, H 3.61%, N 14.15%; Found: C 46.46%, H 3.86%, N 14.67%. TGA shows a 0.7% weight loss between 60-120 °C, which may correspond to absorbed solvent. Decomposition occurs at 175 °C.

**{[Ag(AzoAM*m*P)]$CF_3SO_3$}$_n$ (AgAAM*m*P).** AzoAM*m*P (25.4 mg, 64.4 μmol) in toluene (1.8 mL) was added dropwise to a toluene solution (1.8 mL) of AgOTf (16.6 mg, 64.6 μmol). Upon stirring the reaction mixture for 30 min, an orange-red solid slowly precipitated. $CH_3CN$ (1 mL) was added to re-dissolve the precipitate, and the reaction mixture was stirred at room temperature for 2 h and filtered. Slow evaporation of the filtrate at room temperature provided crystals in orange-red blocks suitable for X-ray analysis. FT-IR (neat, $cm^{-1}$) 3230.2, 3076.0, 2941.6, 2889.8, 2362.8, 2314.7, 2165.8, 1982.2, 1862.7, 1739//.6, 1603.8, 1578.9, 1507.0, 1437.0, 1431.2, 1370.4, 1297.1, 1265.5, 1265.3, 1254.3, 1148.0, 1024.0, 1051.3, 941.1, 741.7, 612.6, 612.3. Elemental analysis calcd. for AgAAMmP $C_{25}H_{24}AgF_3N_6O_4S$: C 44.83%, H 3.58%, N 12.55%; Found: C 44.01%, H 3.31%, N 12.02% The TGA shows no weight loss before decomposition at 177 °C.

**{[Ag(AzoAM*p*P)]$NO_3$}$_n$ (AgAAM*p*P).** AzoAM*p*P (10.0 mg, 25.4 μmol) was dissolved in 2.5 mL $CH_3OH/CH_3CN$ (1:4) mixture with a few drops of DMF. The solution was added dropwise to an $CH_3CN$ solution (0.5 mL) containing silver nitrate ($AgNO_3$, 4.5 mg, 27 μmol) and tetrabutylammonium hexafluorophosphate (*n*-$Bu_4PF_6$, 11.0 mg, 0.0284mmol). The reaction mixture was stirred 30 min to precipitate an orange-red solid. The solid was isolated by filtration, dissolved in 2 mL of a 1:1 mixture of $CH_3OH/CH_3CN$, and for 2 h stirred at room temperature. The mixture was filtered and slow evaporation provided orange-red blocks suitable for X-ray analysis. FT-IR (neat, $cm^{-1}$) 3516.5, 3484.4, 3262.2, 2862.0, 2826.4, 1942.1, 1639.4, 1613.5, 1610.4, 1454.6, 1454.6, 1445.0, 1322.0, 1207.8, 1206.6, 1155.8, 1100.9, 1069.1, 1028.0, 963.3, 889.5, 765.7. Elemental analysis calcd. for AgAAMpP $C_{24}H_{22}N_7O_3Ag$: C 51.03%, H 3.90%, N 17.37%; Found: C 50.92%, H 4.01%, N 17.43%. TGA shows a 3.6% weight loss between 60-120 °C, which may correspond to absorbed solvent. Decomposition occurs at 154 °C.

**[Ag(AzoAE*o*P)]$NO_3$ (AgAAE*o*P).** AzoAE*o*P (30.0 mg, 71.1 μmol) in DCM (1 mL) was added dropwise into an $CH_3CN$ solution (1 mL) of $AgNO_3$ (12.0 mg, 70.6 μmol) and *n*-$Bu_4PF_6$ (28.0 mg, 72.3 μmol). The reaction mixture was stirred 30 min to precipitate an orange solid. The solid was isolated by filtration, dissolved in 2 mL of a 1:1 mixture of $CH_3OH/CH_3CN$, and stirred at room temperature for 2 h. The mixture was filtered and slow evaporation provided orange-red blocks suitable for X-ray analysis. FT-IR (neat, $cm^{-1}$) 3247.2, 3049.8, 2911.3, 2863.4, 2324.7, 2164.5, 2051.2, 1981.6, 1903.6, 1604.1, 1566.1, 1497.9, 1482.5, 1439.3, 1423.7, 1375.0, 1322.0, 1284.7, 1249.4, 1221.3, 1177.2, 1156.4, 1129.0, 1105.9, 1084.8, 1064.8, 1044.1, 1025.4, 1004.8, 959.6, 938.5, 882.3, 846.5, 825.4, 800.5, 762.5, 752.0, 738.4, 647.8, 616.5. Elemental analysis calcd. for AgAAEoP $C_{26}H_{26}N_7O_3Ag$: C 52.67%, H 4.39%, N 16.54%; Found: C 51.84%, H 4.43%, N 16.13%. The TGA shows no weight loss before decomposition at 166 °C.

**{[Ag(AzoAE*p*P)$_2$]PF$_6$}$_n$ (AgAAE*p*P).** AzoAE*p*P (30.0 mg, 71.1 μmol) in DCM (1 mL) was added dropwise into an $CH_3CN$ solution (1 mL) containing $AgNO_3$ (12.0 mg, 70.6 μmol) and *n*-$Bu_4PF_6$ (28.0 mg, 72.3 μmol). The reaction mixture was stirred 30 min to precipitate a yellow solid. The solid was isolated by filtration, dissolved in 2 mL of a 1:1 mixture of $CH_3OH/CH_3CN$, and stirred at room temperature for 2 h. The mixture was filtered and slow evaporation provided orange-red blocks suitable for X-ray analysis. FT-IR (neat, $cm^{-1}$) 3426.5, 3068.1, 2861.8, 2360.3, 2324.7, 2050.9, 1981.4, 1604.0, 1564.4, 1501.5, 1464.6, 1430.7, 1316.2, 1240.8, 1223.5, 1211.6, 1183.9, 1154.6, 1123.0, 1104.1, 1079.5, 1066.5, 1028.6, 939.4, 880.9, 847.3, 823.3, 800.0, 760.1, 740.8, 615.2. Elemental analysis calcd. for AgAAEpP $C_{52}H_{51}N_{12}F_6PAg$: C 56.89%, H 4.65%, N 15.32%; Found: C 56.20%, H 4.28%, N 15.58%. TGA shows a 0.7% weight loss between 60-120°C, which may correspond to absorbed solvent. Decomposition occurs at 197 °C.

**X-ray Crystallography.** Structural analysis was carried out in the X-Ray Crystallographic Facility at Worcester Polytechnic Institute. Crystals were glued on tip of a glass fiber or were covered in PARATONE oil on 100 μm MiTeGen polyimide micromounts and were mounted on a Bruker-AXS APEX CCD diffractometer equipped with an LT-II low temperature device. Diffraction data were collected at room temperature or at 100(2) K using graphite monochromated Mo−Kα radiation (λ = 0.71073 Å) using the omega scan technique. Empirical absorption corrections were applied using the SADABS program.[7] The unit cells and space groups were determined using the SAINT+ program.[7] The structures were solved by direct methods and refined by full matrix least-squares using the SHELXTL program.[8] Refinement was based on $F^2$ using all reflections. All non-hydrogen atoms were refined anisotropically. Hydrogen atoms on carbon atoms were all located in the difference maps and subsequently placed at idealized positions and given isotropic U values 1.2 times that of the carbon atom to which they were bonded. Hydrogen atoms bonded to oxygen atoms were located and refined with isotropic thermal parameters. Mercury 3.1 software was used to examine the molecular structure. Relevant crystallographic information is summarized in Table 1 and Table 2, and the 50% thermal ellipsoid plot is shown in Figures 1-5.

**Powder X-ray diffraction**. PXRD data were collected on a Bruker-AXS D8-Advance diffractometer using Cu-Kα radiation with X-rays generated at 40 kV and 40 mA. Bulk samples of crystals were placed in a 20 cm × 16 cm × 1 mm well in a glass sample holder, and scanned at RT from 3° to 50° (2θ) in 0.05° steps at a scan rate of 2°/min. Simulated PXRD patterns from single crystal data were compared to PXRD patterns of experimental five AzoAX*x*P silver complexes, to confirm the uniformity of the crystalline samples.

**General Spectroscopic Methods.** Solution UV-Vis absorption spectra were acquired in 1.0 cm quartz cuvettes at room temperature and recorded on a Thermo Scientific Evolution 300 UV-Vis spectrometer with inbuilt Cary winUV software. Steady-state diffuse reflectance UV-Vis spectra were obtained on the same instrument with Harrick Praying Mantis diffuse reflectance accessory (Harrick Scientific Products) and referenced to magnesium sulfate. Solution emission spectra were recorded on a Hitachi F-4500 spectrophotometer with excitation and emission slit widths of 5 nm. The excitation source was 150 W Xe arc lamp (Ushio Inc.) operating at a current rate of 5 A and equipped with photomultiplier tube with a power of 400 V.

**Emission and Quantum Yield Determination.** Steady-state emission spectra were recorded on a Hitachi F-4500 spectrophotometer with excitation and emission slit widths of 5 nm. The excitation source was 150 W Xe arc lamp (Ushio Inc.) operating at a current rate of 5 A and equipped with photomultiplier tube with a power of 400 V. Quantum yields in the solid state were determined in triplicated following published procedures using $Na_2SO_4$ as the reference.[9, 10]

**Analyte Detection by Emission.** A 100 µM suspension, with respect to AzoAE*p*P units, of AgAAE*p*P in toluene (2 mL) was prepared, and the emission spectra was recorded ($\lambda_{ex}$ = 523 nm). Upon the addition of each aliquot of analyte, the mixture was equilibrated by stirring for 30 min before recording the emission spectra. The emission response to all analytes was measured by integrating the emission band between 550-800 nm. Pyridine was added from a 2 mM stock solution in toluene in three equal portions, to obtain final concentration of 33, 67, and 100 µM, and the emission was measured. After the third addition, the solvent and pyridine were removed by sparging with $N_2$ gas, and the resulting crystalline material was re-suspended in 2 mL of toluene and stirred for 30 min before re-measuring the emission. The emission loss and restoration steps were repeated in triplicate with single analyte additions to reach 100 µM pyridine. When excess pyridine was added (500 µM), the emission completely disappeared, and could not be restored. Measurements for *N*-methylmorpholine (NMM) were acquired using analogous procedures, were each addition provided a final concentration of 100 µM NMM. Measurements with imidazole were acquired similarly, but emission loss was irreversible. Measurements with dimethylamine (DMA) were acquired similarly except toluene was replaced with THF, and the DMA concentration after each switching cycle was 10 μM. After each DMA addition, the emission was restored by the addition of aliquots of 1.58 mM nitric acid (12.7 μL, 20.1 nmol) to achieve a final $H^+$ concentration of 20 μM.

## RESULTS AND DISCUSSION

*Synthesis and structure*. Using the synthetic protocols for preparing AzoAM*o*P and AzoAE*o*P, we prepared three additional aAB-bipyridyl ligands AzoAM*m*P (**3**, 2,2′-bis[*N,N′*-(3-pyridyl)methyl]diaminoazobenzene), AzoAM*p*P (**4**, 2,2′-bis[*N,N′*-(4-pyridyl)methyl]diaminoazobenzene) and AzoAE*p*P (**5**, 2,2′-bis[*N,N′*-(4-pyridyl)ethyl]diaminoazobenzenee) from the common DAAB scaffold (2,2′-diaminoazobenzene, **6**) building block (Scheme 1). Collectively, we refer to the aABs as AzoAX*x*P compounds where "X" represents the variable components. The *m*-pyridyl ligand containing the ethylene linker (AzoAE*m*P) has not been accessed yet owing to difficulty obtaining a stable and reactive synthon. AzoAE*o*P and AzoAE*p*P are prepared via a Michael reaction with the corresponding vinylpyridine; however, the vinyl group becomes a poor Michael acceptor with the nitrogen atom in the *meta*-position. The anilino nitrogen atom also appears to be a weak nucleophile as reactions with 3-(2-bromoethyl)-pyridine or 2-(pyridin-3-yl)ethyl 4-methylbenzenesulfonate failed to produce the desired product. Attempts to prepare AzoAE*m*P by reductive amination with 2-(pyridin-3-yl)acetaldehyde also proved unsuccessful owing to difficulty isolating sufficient quantities of the aldehyde precursor.

The five ligands can be viewed collectively as a library of extended bipyridine ligands where each pyridine ligand can coordinate a different metal ion to form an extended network.

Fortuitously, all five AzoAX*x*P ligands are highly crystalline in the solid state, and therefore could be subjected to crystallographic analysis. The two pyridine nitrogen atoms of AzoAM*o*P are 8.37 Å apart; however, the pyridine lone pairs engage in hydrogen bonding with the anilino hydrogen atom, which orients the potential coordination sites inward. In preliminary screenings, we observed that AzoAM*o*P did not appear to coordinate metal ions like $Zn^{2+}$, presumably because the thermodynamic stability of the hydrogen bonds. We reasoned that a thermodynamically stable, kinetically inert metal-ligand bond could provide access to a metal complex with AzoAM*o*P by shifting the equilibrium from hydrogen bonding to metal ligand bonding toward complex formation. $Ag^{+}$ appeared to be a good candidate for coordination polymer formation based on other successful investigations,[11-15] as well as the stability of Ag–pyridine bonds.[16] We further speculated that the photoactivity of $Ag^{+}$ might provide access to additional functionality in the resulting AB materials.

The other four AzoAX*x*P derivatives do not exhibit hydrogen bonding with the pyridine lone pairs, so ostensibly the ligands could bind metal ions more like typical bridging bipyridine derivatives. Two-coordinate metals with bridging bipyridine ligands are rare[17-19] except for $Ag^{+}$.[11, 13, 20, 21] The two AzoAX*x*P derivatives containing methylene linkers, AzoAM*m*P and AzoAM*p*P, can rotate at the azobenzene–anilino C–N bond, anilino–methlyene N–C bond and methylene–pyridine C–C bond. The two AzoAX*x*Ps with ethylene linkers, AzoAE*o*P and AzoAE*p*P, can also rotate about C–C ethylene bond. The multiple degrees of freedom suggest our ligands might go through noticeable reorientation during the formation of $Ag^{+}$ complexes or polymers.

Although not absolutely predictive, the geometric structure of $Ag^{+}$ metal organic polymer chains depends on the pyridine substitution pattern, the directionality of the pyridine lone pairs, and the conformational freedom of the ligand. Solvent and anion interactions also can impact the extended macromolecular structure. In bipyridine ligands containing *ortho* pyridine nitrogen atoms, the lone pairs tend to face inward in the apo form, but often reorient during polymerization process to minimize crowding between the side chain and the $Ag^+$ binding site, which often results in zigzag silver chain structures.[21, 22] In *meta*- and *para*-pyridine bipyridine ligands, the lone pairs tend to face outward, and therefore do not necessarily need to undergo structural rearrangement in two-coordinate $Ag^+$ structures. Some *meta*-bipyridine derivatives exhibit minimal reorientation,[23] or asymmetric changes on one half of the molecule.[21] In more rigid *meta*- and *para*-bipyridine ligands, linear geometries are more common because of partly or completely restricted movement.[24, 25]

Based on the structural trends with other bipy derivatives,[21, 22] we hypothesized that the pyridine nitrogen lone pairs in AzoAM*o*P would flip outward to accommodate $Ag^+$ coordination if the intramolecular hydrogen bonds were disrupted. We prepared the AzoAM*o*P complex using protocols designed to gradually deliver $Ag^+$ through ligand exchange with $CH_3CN$.[13] A $[Ag(AzoAMoP)]^+$ complex precipitates immediately from toluene solution, but re-dissolves upon addition of $CH_3CN$. Slow evaporation of $CH_3CN$ leads to nitrile–pyridine ligand exchange, and the formation of a AgAAM*o*P coordination polymer, which was isolated as well-defined orange crystals. Compared to the structure of the apo ligand, the azo N–C and anilino–methylene N–C bonds in the $Ag^+$ polymer with AzoAM*o*P ligand are rotated by 180˚ and 81˚ respectively (Figure 1). This rearrangement orients the two pyridine rings perpendicular to the azo core but retains inward-facing pyridine nitrogen lone pairs in the formation of a helical structure where each $Ag^+$ is coordinated by a pyridine nitrogen atoms from each of two different AzoAM*o*P ligands. $Ag^+$ binding increases the length of the hydrogen bonds between the anilino hydrogen atoms and the diazene lone pairs of DAAB core by approximately 0.4 Å, and introduces a slight asymmetry in the two halves of the molecule. The core N=N bond decreases slightly from 1.28 Å to 1.26 Å owing to the decrease in hydrogen bonding (Table 1). Although the changes in hydrogen bonding might suggest a less rigid and therefore less emissive azobenzene chromophore, we expected that the metal complexation and crystal packing interactions would offset the loss of intramolecular forces.

Compared to AzoAM*o*P, AzoAM*m*P and AzoAM*p*P exhibit more conformational freedom, since both lack hydrogen bonding between the anilino hydrogen atoms and pyridine nitrogen atoms. The point-to-point distance between the two pyridine nitrogen atoms distances are 15.96 Å and 16.52 Å respectively, but unlike linear bridging analog like 4,4'-bipyridine, the effective distance obtained by projecting one pyridine nitrogen on a plane containing the other pyridine is shorter, 14.76 Å and 14.70 Å, respectively. We calculate the effective distance by combining the offset between the two pyridine ligands in the y and z direction where the x-axis is defined by a line going through one pyridine nitrogen atom through its para carbon atom, and defining the xy plane by that pyridine ring. The y and z offset distances are therefore the displacement from a linear bipyridine ligand (e.g. 4,4'-bipyridine) from the defined x-axis. So for AzoAM*m*P the pyridine nitrogen atoms are offset from linearity by 5.41 Å (y) and 2.12 Å (z, Figure S33), and AzoAM*p*P has more displacement along the y trajectory (6.70 Å), but a shorter deviation in the z direction (1.79 Å, Figure S34). In contrast to AzoAM*o*P, the pyridine lone pairs in both AzoAM*m*P and AzoAM*p*P point away from the diazene core to afford extended bipyridine derivatives more reminiscent of the bridging ligands found in coordination polymers. Structural changes in *meta*- and *para*-bipy derivatives are typically minimal during $Ag^+$ complex formation due to limited or no orientational freedom.[21, 23-25] Based on these trends, we expected the AgAAM*m*P and AgAAM*p*P polymers, where the pyridine nitrogen atoms face outward in the apo ligands, would likewise undergo minimal structural changes.

AgAAM*m*P was synthesized by identical procedures used to prepare AgAAM*o*P. Due to limited solubility of AzoAM*p*P ligand in toluene however, the solvent was replaced with a 1:4 mixture of $CH_3OH/CH_3CN$ and a minimal amount of DMF, and AgOTf was substituted with $AgNO_3$. To obtain x-ray quality crystals, anion metathesis with *n*-$Bu_4PF_6$ was used to replace the $NO_3^-$ anion with $PF_6^-$. Compared to the structure of the apo ligand, the azo N–C bonds in the $Ag^+$ polymer with AzoAM*m*P and AzoAM*p*P ligands are rotated by 180° (Figure 2 and Figure 3), which is identical to the azo N–C bond rotation observed during the formation of AgAAM*o*P. This reorients the two pyridine rings, but retains outward-facing pyridine nitrogen lone pairs in the formation of a linear or zigzag chain where each $Ag^+$ is coordinated by a pyridine nitrogen atoms from each of two different AzoAM*m*P or AzoAM*p*P ligands.

The DAAB core in AzoAM*m*P and AzoAM*p*P exhibit nearly identical the hydrogen bond lengths, 2.33 and 2.34 Å respectively; however $Ag^+$ binding produces different changes in the three derivatives containing a methylene spacer. Unlike AgAAM*o*P, the hydrogen bond length between the anilino hydrogen atoms and the diazene lone pairs decreases. While the hydrogen bonds in AgAAM*p*P (2.02 Å) are symmetric, AgAAM*m*P contains a longer (2.12 Å) and shorter (1.96 Å) bond that are both contracted compared to the apo-ligand. While AgAAM*o*P exhibited some asymmetry in the hydrogen bond length, the difference is much more pronounced in AgAAM*m*P. As expected, the increased hydrogen bonding interaction results in the opposite effect on the N=N bond lengths, which increases slightly from 1.27 Å to 1.28 Å for both coordination polymers.

The pyridine nitrogen atoms of AzoAE*o*P are separated by 15.04 Å. While the pyridine lone pairs face inward, the length of the ethylene linkers and ligand flexibility provide no a priori reason to predict a reorientation would be required to coordinate $Ag^+$. Using the identical procedure used to prepare AgAAM*p*P, a discrete monomeric complex was obtained (Figure 4). The single rotation about the azo C–N required to achieve this coordination geometry suggests kinetic trapping of a 17-membered metallacycle instead of a polymer. To the best of our knowledge, this is the largest Ag metallacycle formed from a bipy derivative to date. Large metallacycles with pyridyl donor groups are relatively uncommon. Most analogously, 13-membered-ring metallacycles of Pt and Pd,[26] and an 11-membered-ring metallacycle of Ag have been observed.[22]

The intramolecular hydrogen bonds in AzoAE*o*P may be a significant factor in the preference for metallacycle formation instead of a polymeric structure. Two 2.36 Å hydrogen bonds between anilino hydrogen atom and the azo nitrogen atoms appear in apo ligand. When the ligand rotates to close the metallacycle, the hydrogen bonding between one of the anilino hydrogen and azo nitrogen is unavailable, and a new 2.07 Å hydrogen bond forms between the same proton and the other azo nitrogen atom. The hydrogen bonding in the upper half molecule remains unchanged, but the distance increases slightly from 2.36 to 2.38 Å. The shorter, stronger hydrogen bond formed in AgAAE*o*P may provide a thermodynamic driving force for metallacycle formation.

Of the five AzoAX*x*P ligands, AzoAE*p*P most closely resembles a linear bridging ligand like 4,4'-bipyridine. The point-to-point distance between the pyridine nitrogen atoms of 17.95 Å is the longest distance of all the AzoAX*x*P ligands, but the trajectory of the two pyridine groups is linear and the aromatic rings are coplanar in the solid state. Linear bipyridines with long nitrogen-nitrogen distances form both linear $Ag^+$ polymers[25, 27] and three-dimensional networks with 4-coordinate $Ag^+$ sites.[28, 29] The preference for 2-coordinate or 4-coordinate $Ag^+$ appears to correlate with the reorientation ability and rigidity of the ligands as well as the distance between pyridine nitrogen atoms.

Although the pyridine ligands in apo-AzoAE*p*P are coplanar, rotation about the C–N bond in the ethylene spacers results decreases the point-to-point distance from 17.95 Å to 6.65 Å. The effective distance is obtained similarly as described above to yield y and z offset distances of 2.22 and 9.09 Å, respectively (Figure S35). Each $Ag^+$ is four-coordinate with each pyridine coming from one of four unique AzoAE*p*P ligands. We partially attribute the increased coordination to the reduced steric requirements at the donor ligand site of the *para*-pyridine isomer. Each AzoAE*p*P ligand also coordinates two separate Ag cations to form a three-dimensional infinite polymer with the distorted tetrahedral $Ag^+$ sites (Figure 5B). The lack of steric hindrance at the coordination site of the *para*-pyridine ligands appears to be an important prerequisite for three-dimensional polymer formation with bipyridines. In previous $Ag^+$ polymers and three-dimensional networks with 4-coordinate $Ag^+$ sites, most of the ligands undergo minimum or no reorientation.[25, 27, 29] Unlike these linear bipyridine ligands, AzoAE*p*P exhibits a high degree of freedom, which allows optimization of steric and electrostatic factors in $Ag^+$ complex formation.

*Emission.* AzoAM*o*P exhibits no measurable emission at room temperature, but emits when frozen in a solvent glass at 77 K. The four additional AB compounds behave similarly. Although the wavelengths vary, all the five compounds emit with a $\lambda_{max}$ between 566 nm and 610 nm (Table 2). While quantitative measurements of minimally emissive complexes are imprecise, qualitatively, AzoAE*o*P and AzoAE*p*P exhibit brighter emission than AzoAM*m*P and AzoAM*p*P. AzoAM*o*P appears to emit more weakly than the other four derivatives.

To assess the photochemistry of the $Ag^+$ complexes, the emission of AgAAM*o*P, AgAAM*m*P, AgAAM*p*P, and AgAAE*p*P were evaluated in toluene as suspensions of powdered crystals. The four coordination polymers form semi-homogeneous dispersions in toluene that remain suspended for several months. Similar to the free ligands, the emission $\lambda_{max}$ occurs around 600 nm for all the complexes (Table 2). In the solid state, the relative integrated emission intensities reveal a nearly 30-fold brighter emission for the four-coordinate AgAAE*p*P complex compared to the least emissive AgAAM*p*P complex (Figure 6). Complex formation redshifts the absorbance maximum of AzoAE*p*P from 502 nm to 594 nm with an emission peak centered at 641 nm (Figure 7). Similar to the absorbance, the emission of AgAAE*p*P redshifts 31 nm from that of free AzoAE*p*P measured at 77 K, which is a greater a change than the other three polymeric complexes. The luminescence response appears to validate the initial hypothesis that embedding the AzoAX*x*P ligand in a solid state material increases the degree of radiative decay of the AB excited state.

The emission of AB in solution at room temperature is too weak to be detected by conventional emission instruments because efficient non-radiative pathways for the decay of the excited state exist.[30] In many aromatic chromophores, the emission of a photon accompanies a $S_1(\pi\pi^*)\rightarrow S_0$ transition as the excited molecule returns to the ground state. In AB however, the $S_2(\pi\pi^*)\leftarrow S_0$ absorbance has the largest extinction coefficient, but a weakly absorbing interstitial $S_1(n\pi^*)$ state is present. While the photophysical processes remain an active area of investigation,[31-33] a rapid intersystem crossing to a vibrationally excited $S_1(n\pi^*)$ state occurs after $S_2(\pi\pi^*)\leftarrow S_0$ absorbance. AB isomerization occurs via the concerted inversion pathway from the $S_1$ state.[34, 35] A modest increase in emission intensity can be observed in a frozen matrix when relaxation pathways involving molecular motion are impaired.

Radiative decay of AB can be increased by manipulating the frontier orbitals of the chromophore as observed in 2-borylazobenzenes.[36-39] Engaging the diazene lone pair in a dative bond with a boron atom lowers the energy of the lone pair, or B–N bonding orbital, below the diazene π–bond. Thus forming the nπ* state would require promoting an electron into the half-occupied, higher energy diazene π-orbital. The absence of the interstitial $S_1$(nπ*) state removes the optically forbidden $S_1$(nπ*)→$S_0$ transition and opens the $S_2$(ππ*)→$S_0$ emission channel.

We hypothesized that introducing intramolecular hydrogen bonds between the diazene lone pair and anilio hydrogen atom might lead to a similar, albeit less drastic, change in the frontier orbitals to those observed in 2-borylazobenzenes, and therefore increase radiative decay of the excited state. To support this hypothesis, we interrogated the ground state electronic structure of each ligand using Density functional theory (DFT). In order to capture the structural impact of the solid-state environment on the electronic structure for each ligand, unit cell vectors and atom positions were determined using three-dimensional periodic boundary conditions before extracting a single molecule for analysing the electronic structure, an approach we have previously applied to luminescence metal-organic systems.[40] The first ten excited states were computed vertically using time-dependent DFT (TDDFT) for all the ligands.

The computational approach provides values that correlate closely with experimentally measured absorption energies. As expected, the gap between the n and π* orbitals decreases significantly going from AB to DAAB, but the order remains unchanged (Figure S36). The red-shifting of the ππ* absorption maxima between AB and DAAB was calculated to be 0.95 eV, in reasonable agreement with the experimental value of 1.23 eV. Likewise, the calculated and experimental values for the five aAB showed reasonable agreement. The minimum and maximum red-shifting of the ππ* transition of the five aAB ligands were measured as 0.139 eV and 0.142 eV with respect to DAAB. The equivalent red-shifts determined computationally were 0.250 eV and 0.294 eV, a similar spread to the experimental values shifted by approximately 0.1 eV.

Since our series of modified aAB ligands are significantly more emissive in a frozen matrix than AB and DAAB, we reasoned that the anilino substituent must induce further changes in the frontier orbitals. In examining the frontier orbital energies (Table S3), we observed that the pyridine moieties induce a stabilization of the diazene-centered n-type orbital for the four aAB ligands (maximum 0.33 eV). This observation is in line with the expectation that the inductive effect of the pyridine substituent results in a slightly more electron rich secondary anilino nitrogen atom, leading to a slightly more stable $\pi^*$ orbital and making the diazene lone pairs somewhat better hydrogen bond acceptors. AzoAM*o*P, which exhibits a relative orbital destabilization of approximately 0.44 eV, provides an exception to this trend. AzoAM*o*P shows a reduced gap between the n and $\pi^*$ orbitals compared to the other four derivatives. We attribute this difference to the additional hydrogen bonding provided by the pyridine substituent, reducing the interaction between the diazene lone pair and the anilino hydrogen atom. In contrast, introducing a *meta*/*para* nitrogen atom within the pyridine unit or an ethylene spacer increases the distance between the nitrogen donor atom and the diazene core, rendering this interaction negligible and yielding a net stabilization of the lone pair, as is seen in the other four derivatives. The lack of emission in solution, and subsequent increase in a frozen matrix, is also consistent with hydrogen bonding as a key contributor to the emission behavior. The restricted motion at low temperature would be expected to lead to a stabilization of the hydrogen bonds, and, in turn, the energy levels of the frontier orbitals, that would not necessarily occur in solution.

The electron donating pyridine moieties destabilize the π **(**HOMO) orbital for all ligands with respect to the DAAB control molecule. This destabilization exceeds the stabilization of the diazene lone pair orbital (LUMO), which results from hydrogen bonding between the diazene lone pair and the anilino hydrogen atom (maximum 0.95 eV). AzoAM*o*P again provides an exception, where the π* orbitals are stabilized upon the addition of the pyridine substituents (0.52 eV). The combination of these effects results in a general stabilization of the ππ* excited state and destabilization of the nπ* state (Figure S36), causing an inversion of the emissive ππ* and non-emissive nπ* states with respect to both AB and DAAB. As expected, this results in the removal of the lower-lying nπ* state as a radiationless decay channel.[39] Notably, the inversion of states is the least pronounced in AzoAM*o*P. In an absolute sense, the nπ* state still lies higher in energy than the ππ*, but the small energy gap (0.04 eV) could permit internal conversion between these two states, accounting for the lower emission intensity of AzoAM*o*P.

To generate emission at room temperature, we hypothesized that embedding the ligands in coordination polymers would provide the requisite restrictions on molecular motion needed to stabilize the diazene lone pair-anilino hydrogen atom hydrogen bond. Our hypothesis was validated by the emission measurements on the $Ag^+$ systems. The 2.43 Å diazene lone pair-anilino hydrogen atom bond length in $[Ag(AzoAM\textit{o}P)]_n$ is the longest of the four polymeric complexes, and correlates with the lowest relative integrated emission intensity (Table 2). As the hydrogen bond length shortens in $[Ag(AzoAM\textit{m}P)]_n$ (2.12 Å) and $[Ag(AzoAM\textit{p}P)]_n$ (2.03 Å), the relative emission intensity increases. As our hypothesis predicts, $[Ag(AzoAE\textit{p}P)_2]_n$ has the brightest emission and the shortest hydrogen bonds. In order to understand the effects of the structural reorganization imposed by $Ag^+$ coordination on the frontier orbitals and excited states, the same computation procedures were applied to the polymeric complexes. These calculations only capture the structural effects of the $Ag^+$ coordination since a neutral ligand is extracted from the periodic, solid state calculations; therefore, the model does not account for the electronic effects of $Ag^+$ coordination.

While the trend between hydrogen bond length and emission intensity correlates as expected, the computational results predict a small energy gap between the n and π* orbitals. Since the structural reorganization of the anilino moieties after polymer formation causes only a slight modulation in the lowest-lying excited states (Figure S37), the qualitative ordering of the states remains unchanged with respect to the results obtained for the crystalline ligands alone. Interestingly, $[Ag(AzoAE\textit{p}P)_2]_n$ is over an order of magnitude more emissive than the other three polymeric complexes, which indicates other contributing factors to the photophysical behavior. The calculations based on extracted ligands only capture structural effects imposed by coordination to $Ag^+$; therefore, while we are able to demonstrate a correlation between hydrogen bond length and emission behavior across the series of materials, we cannot draw any conclusions regarding the influence of $Ag^+$ atoms or the crystalline environment on orbital or excited state energies. Together, these results show that the enhanced luminescence observed in these azobenzene derivatives could be based on the inversion of the aforementioned electronic excited states relative to azobenzene, but no conclusions can be made concerning the differences in emission intensity observed upon the formation of silver complexes, or the differences in luminescence intensity observed between different silver-based polymers.

AgAAE*p*P exhibits the brightest emission of all the $Ag^+$ coordination polymers with a quantum yield of 0.8%. We examined the emission response to the $Ag^+$-coordinating analytes pyridine, *N*-methylmorpholine, DMA and imidazole. When 5 equivalents of pyridine with respect to AzoAE*p*P units is added to a suspension of AgAAE*p*P in toluene, no emission was detected; however, the absorbance spectrum showed the presence of AzoAE*p*P. When a sub-stoichiometric amount of pyridine was added, the coordination polymer emission decreased. We suspected that pyridine was displacing AzoAE*p*P pyridine ligands bound to the $Ag^+$ sites, which would cause the loss of extended structure in the coordination polymer. The release of AzoAE*p*P, which is not emissive at room temperature in solution, would produce an on/off switching behavior. To provide supporting evidence for this signaling mechanism, the pyridine was removed by sparging the solution with $N_2$ gas. The recovery of emission after the evaporation and re-dispersion of the material in toluene suggests the reassembly of the original AgAAE*p*P structure. This reversible process can be monitored by emission spectroscopy. Initially, the emission spectrum of AgAAE*p*P exhibits features with maxima at 620 nm (Figure 8A). Upon the addition of pyridine, the peak intensity decreases gradually, and correlates with expected changes in absorbance. The pyridine removal process restores the original spectrum characteristic of AgAAE*p*P. This process is fairly robust, and can be repeated with minimal loss of emission. After five cycles, the material retains 90% of the original emission intensity (Figure 8B). NMM shows similar results to pyridine. The boiling points of pyridine and NMM are 115.2 °C and 116 °C, respectively, so the sparging with $N_2$ removes both the analyte and solvent (toluene, bp 111°C).

In contrast, when exposed to imidazole, the emission loss is not reversible. Although imidazole displaces AzoAE*p*P from the coordination sphere of $Ag^+$ in the coordination polymer, the lack of volatility (bp 256 °C) makes removal nearly impossible. A similar response occurs upon treatment of AgAAE*p*P with potassium bromide. The removal of $Ag^+$ through the formation of insoluble halide complexes leads to an irreversible loss of emission and a UV spectrum consistent with free AzoAE*p*P.

The ability to reversibly detect analytes through changes in extended structure is not limited to removal by evaporation. Addition of DMA leads to a loss of emission analogous to the response to pyridine. The addition of dilute nitric acid causes the AgAAE*p*P to reassemble. This process can also be repeated for multiple cycles without significant loss of absolute emission intensity (Figure 8C). The emission loss and restoration process by adding and removing volatile analytes like pyridine and NMM demonstrates the viability of a disassembly-reassembly process for emission detection. While physical/electrostatic interactions are more common signal transduction pathways,[41-43] our system provides an alternative sensing mechanism driven by coordination events.

## CONCLUSION

Although ABs are not typically emissive, embedding AB chromophores in a rigid coordination polymer can enhance the degree of radiative decay of the excited state. In the aAB ligands investigated, both $Ag^+$ coordination and intramolecular hydrogen bonds contribute to the emission enhancement; however, the hydrogen bonding introduces the majority of the change in the frontier molecular orbitals. The stabilization of the ππ* excited state below the interstitial nπ* usually involved in the non-radiative decay of the excited state leads to luminescence in both solid state and crystalline samples dispersed in solvent. Our initial investigations suggest three-dimensional networks are more emissive than linear coordination polymers, however, additional examples will be required to confirm this hypothesis. The coordination polymers can be completely, or partially disassembled by the addition of $Ag^+$-binding analytes such as amines, and reformed after analyte removal to provide a sensor-like system. This signal transduction mechanism differs from many polymeric sensors, and will be investigated as an alternative approach to designing practical luminescent probes.

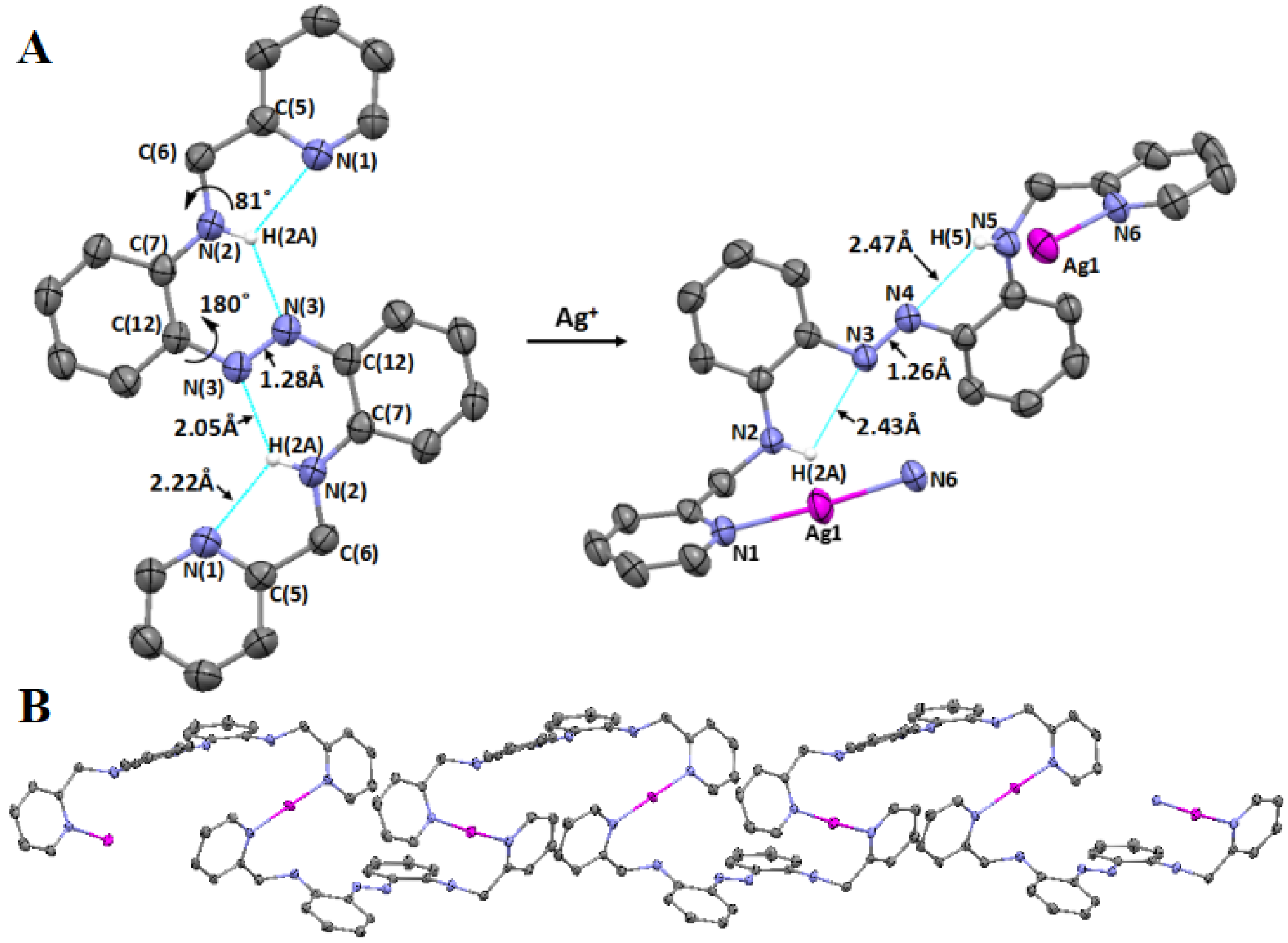

**Figure 1A.** Preparation of AgAAM*o*P showing the ORTEP diagram of AzoAM*o*P and AgAAM*o*P with 50% thermal ellipsoids and selected atom labels. Hydrogen atoms except for those engaged in intramolecular hydrogen bonds are omitted for clarity. The triflate anion and a non-coordinating $CH_3CN$ solvent molecule are omitted for clarity. AgAAM*o*P was prepared in toluene using AgOTf as the silver source using a slow ligand exchange process with $CH_3CN$. During the polymer formation, the azo N–C and anilino–methylene N–C bonds rotate by 180˚ and 81˚, respectively. **1B.** ORTEP diagram showing 50% thermal ellipsoids of the expanded helical polymer chain. Hydrogen atoms, triflate anions and CH3CN solvent molecules are omitted for clarity.

A

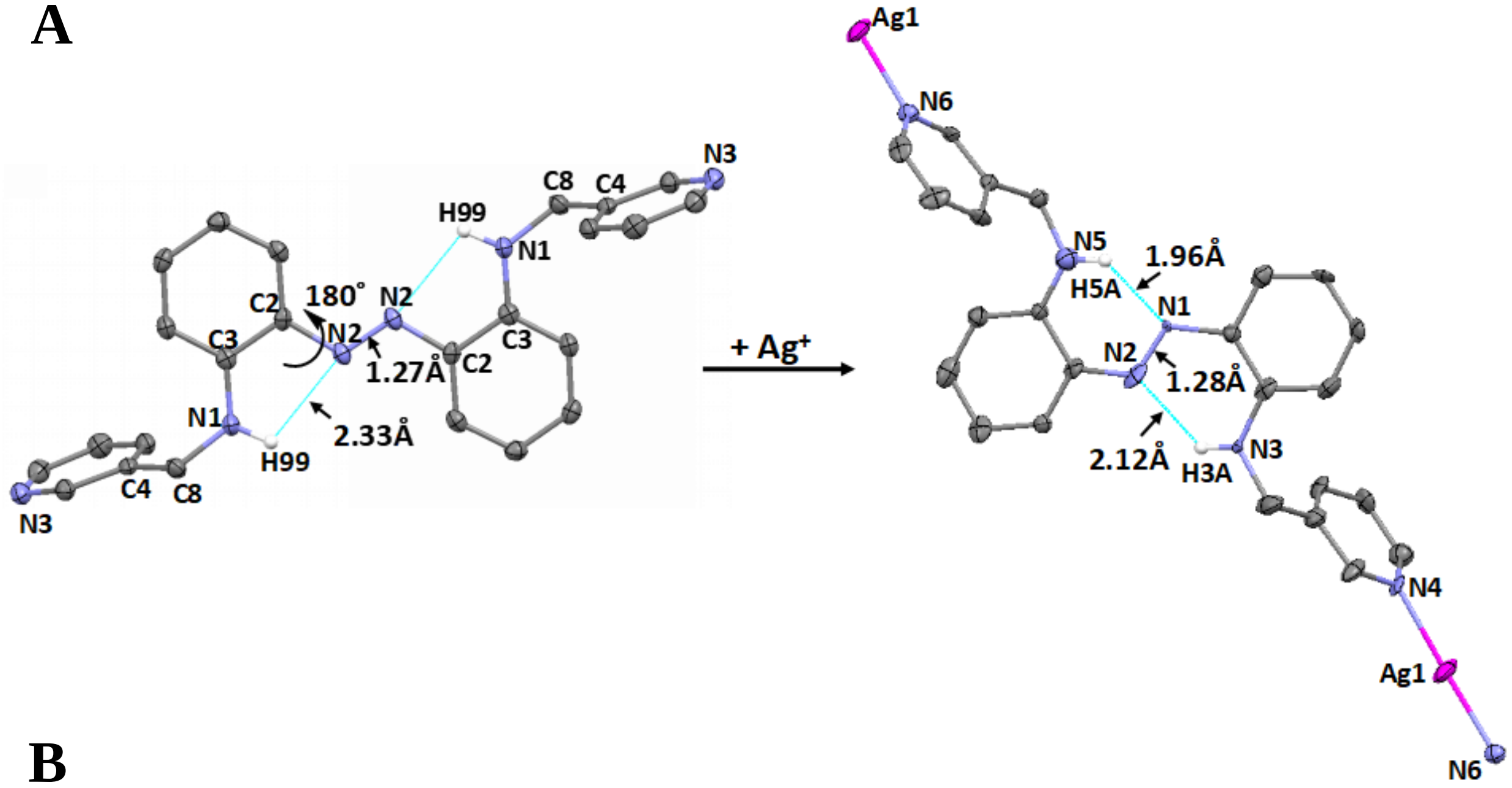

B

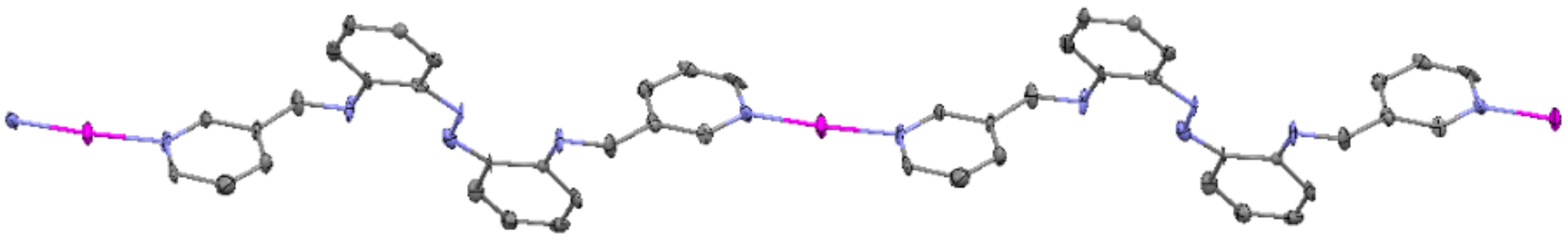

**Figure 2A**. Preparation of AgAAM*m*P showing the ORTEP diagram of AzoAM*m*P and AgAAM*m*P with 50% thermal ellipsoids and selected atom labels. Hydrogen atoms except for those engaged in intramolecular hydrogen bonds are omitted for clarity. The triflate anion is removed for clarity. AgAAM*m*P was prepared in toluene using AgOTf as the silver source using a slow ligand exchange process with $CH_3CN$. During the polymer formation, the azo N–C bonds rotate by 180˚. **2B.** ORTEP diagram showing 50% thermal ellipsoids of the expanded helical polymer chain. Hydrogen atoms, triflate anions are omitted for clarity.

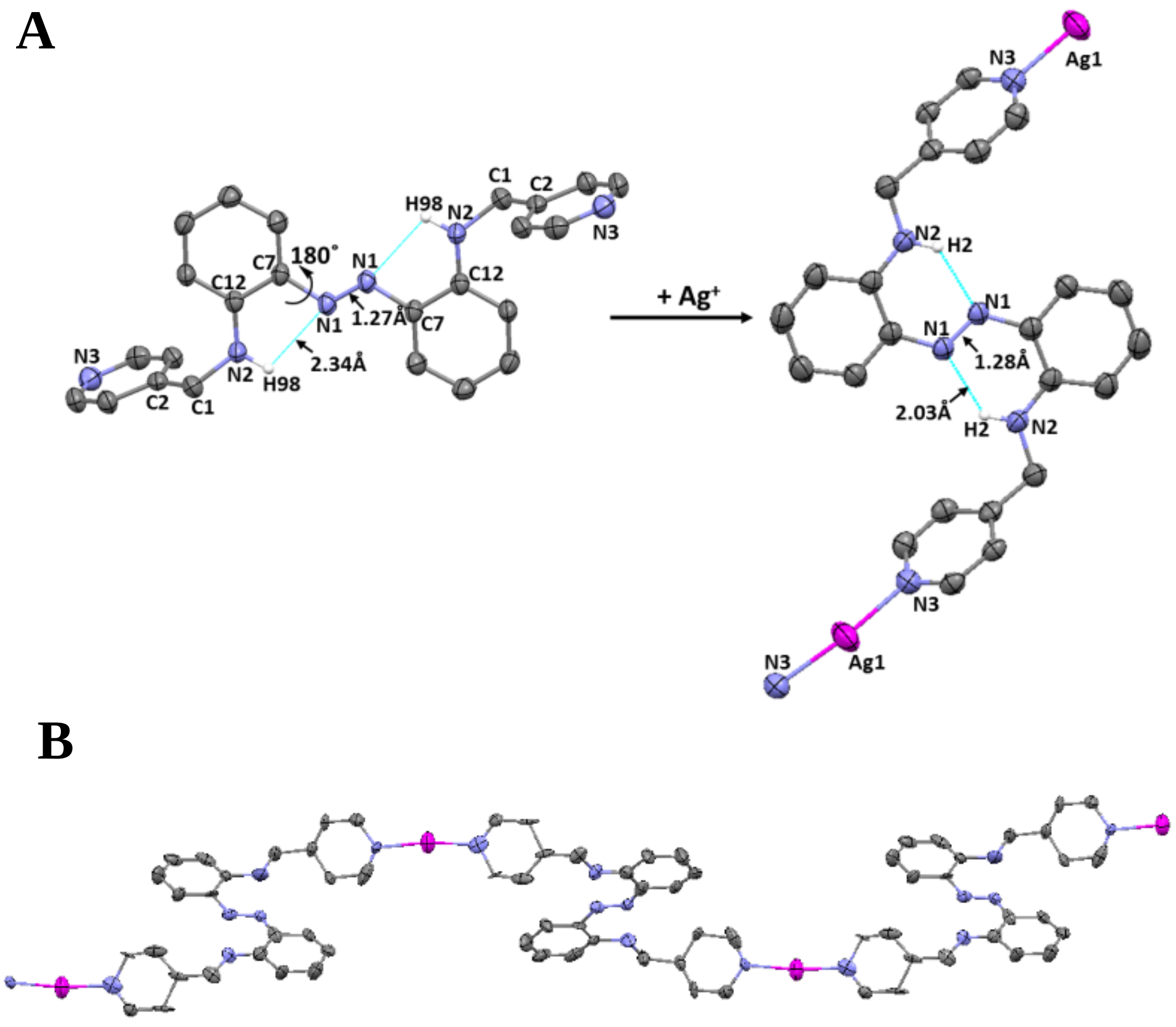

**Figure 3A**. Preparation of AgAAM*p*P showing the ORTEP diagram of AzoAM*p*P and AgAAM*p*P with 50% thermal ellipsoids and selected atom labels. Hydrogen atoms except for those engaged in intramolecular hydrogen bonds are omitted for clarity. The $PF_6$ anion is removed for clarity. AgAAM*p*P was prepared in $CH_3OH/CH_3CN$ (1:4) using $AgNO_3$ as the silver source. Subsequent anion exchange with *n*-$Bu_4PF_6$ in $CH_3CN$ provided x-ray quality crystals. During the polymer formation, the azo N–C bonds rotate by 180˚. **3B.** ORTEP diagram showing 50% thermal ellipsoids of the expanded zigzag polymer chain. Hydrogen atoms, $PF_6$ anions are omitted for clarity.

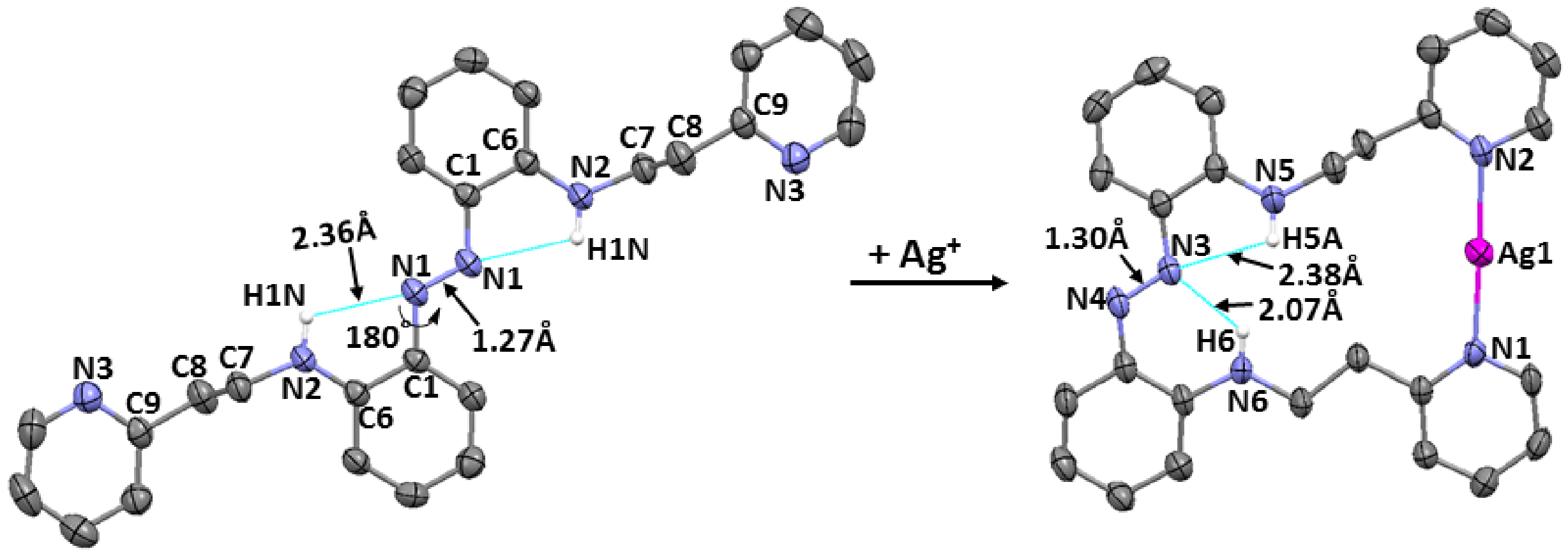

**Figure 4.** Preparation of AgAAE*o*P showing the ORTEP diagram of AzoAE*o*P and AgAAE*o*P with 50% thermal ellipsoids and selected atom labels. Hydrogen atoms except for those engaged in intramolecular hydrogen bonds are omitted for clarity. The $PF_6$ anion is removed for clarity. AgAAE*o*P was prepared in $CH_3OH/CH_3CN$ (1:4) using $AgNO_3$ as the silver source with added *n*-$Bu_4PF_6$ to provide x-ray quality crystals. During the complex formation, one azo N–C bond rotates by 180° while the other remains stationary to form a 17-membered ring. The two hydrogen bonds formed between two anilino hydrogen atoms and the same azo N atom are asymmetric with lengths of 2.07 Å and 2.38 Å respectively.

A

B

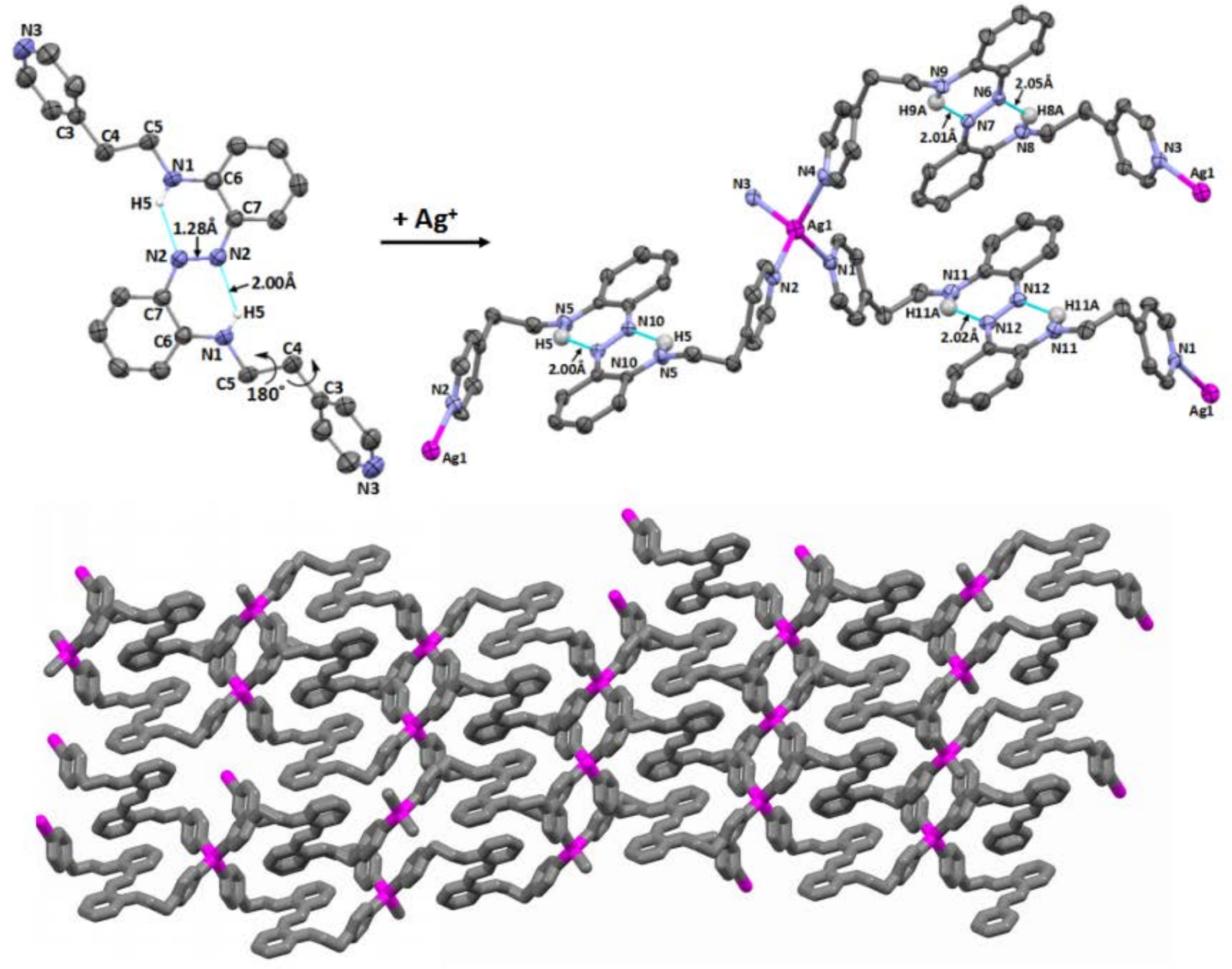

**Figure 5A**. Preparation of AgAAE*p*P showing the ORTEP diagram of AzoAE*p*P and AgAAE*p*P with 50% thermal ellipsoids and selected atom labels. Hydrogen atoms except for those engaged in intramolecular hydrogen bonds are omitted for clarity. The $PF_6$ anion is removed for clarity. AgAAE*p*P was prepared in DCM using $AgNO_3$ as the silver source with added *n*-$Bu_4PF_6$. The resulting precipitate was dissolved in $CH_3CN$ to yield x-ray quality crystals from slow solvent evaporation. During the polymer formation, the ethylene C–C bonds rotate by 180˚ and ethylene-pyridine C–C rotates minimally. **5B.** ORTEP diagram showing 50% thermal ellipsoids of the three dimensional polymer chain. Hydrogen atoms and $PF_6$ anions are omitted for clarity.

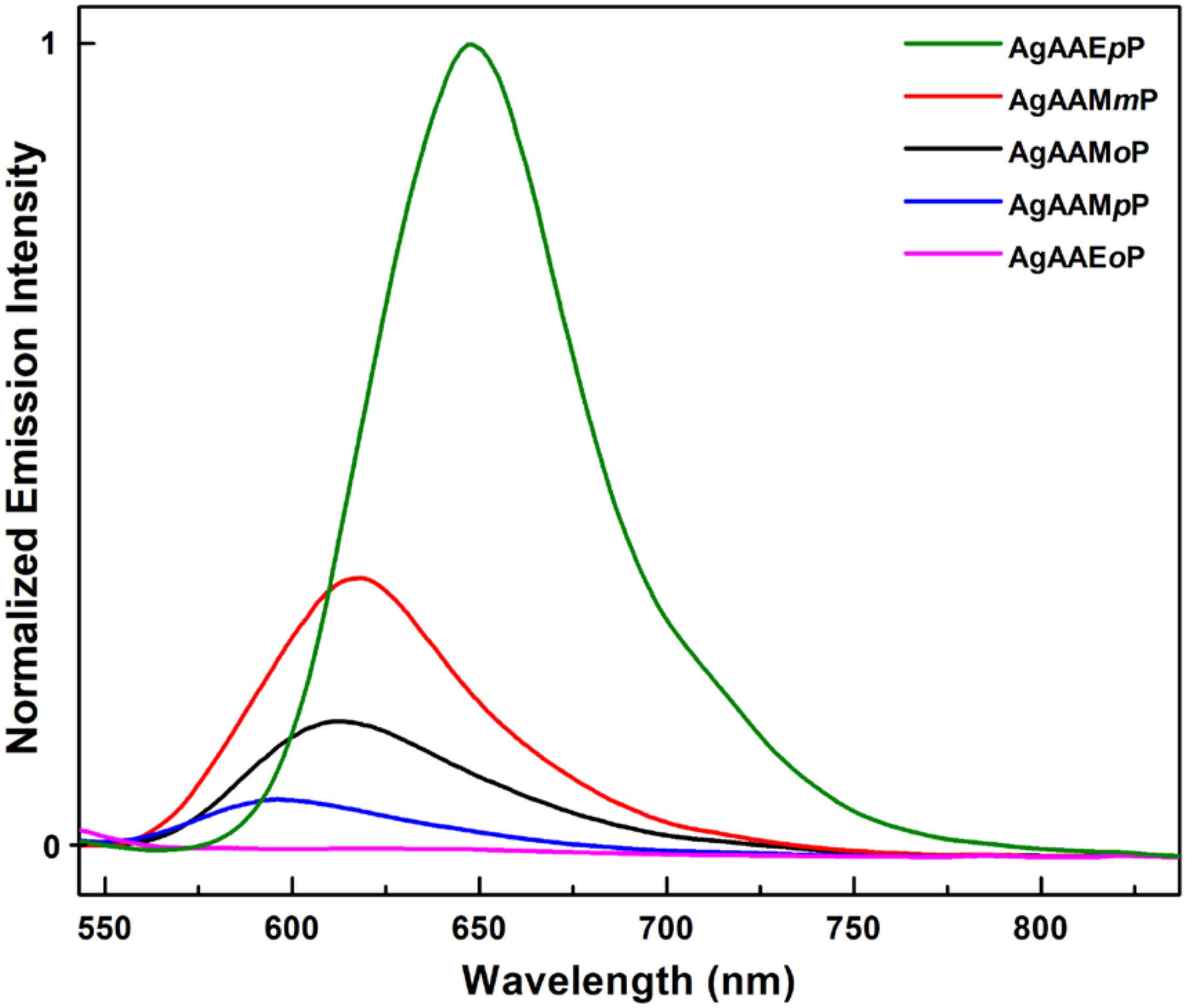

**Figure 6.** Solid state emission spectrum of the five AzoAX*x*P silver complexes ($\lambda_{ex}$ 523 nm) showing the relative intensities.

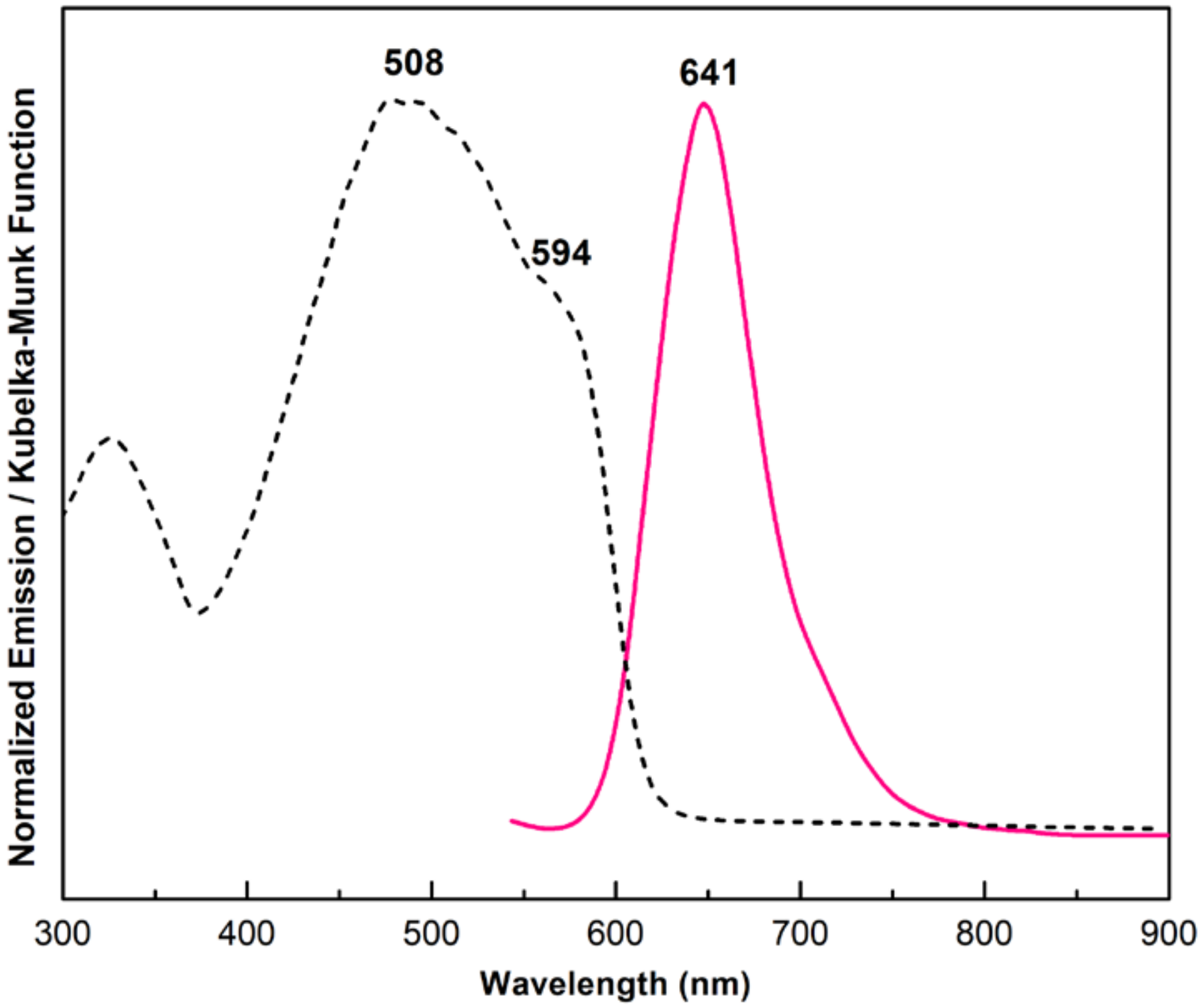

**Figure 7.** Solid state diffuse reflectance (black) and solid state emission ($\lambda_{ex}$ = 523 nm, pink) spectra of AgAAE*p*P. Inset: cuvettes containing a suspension of the complex crystals in toluene (100 μM) without irradiation (left), and excited with 365 nm light (right).

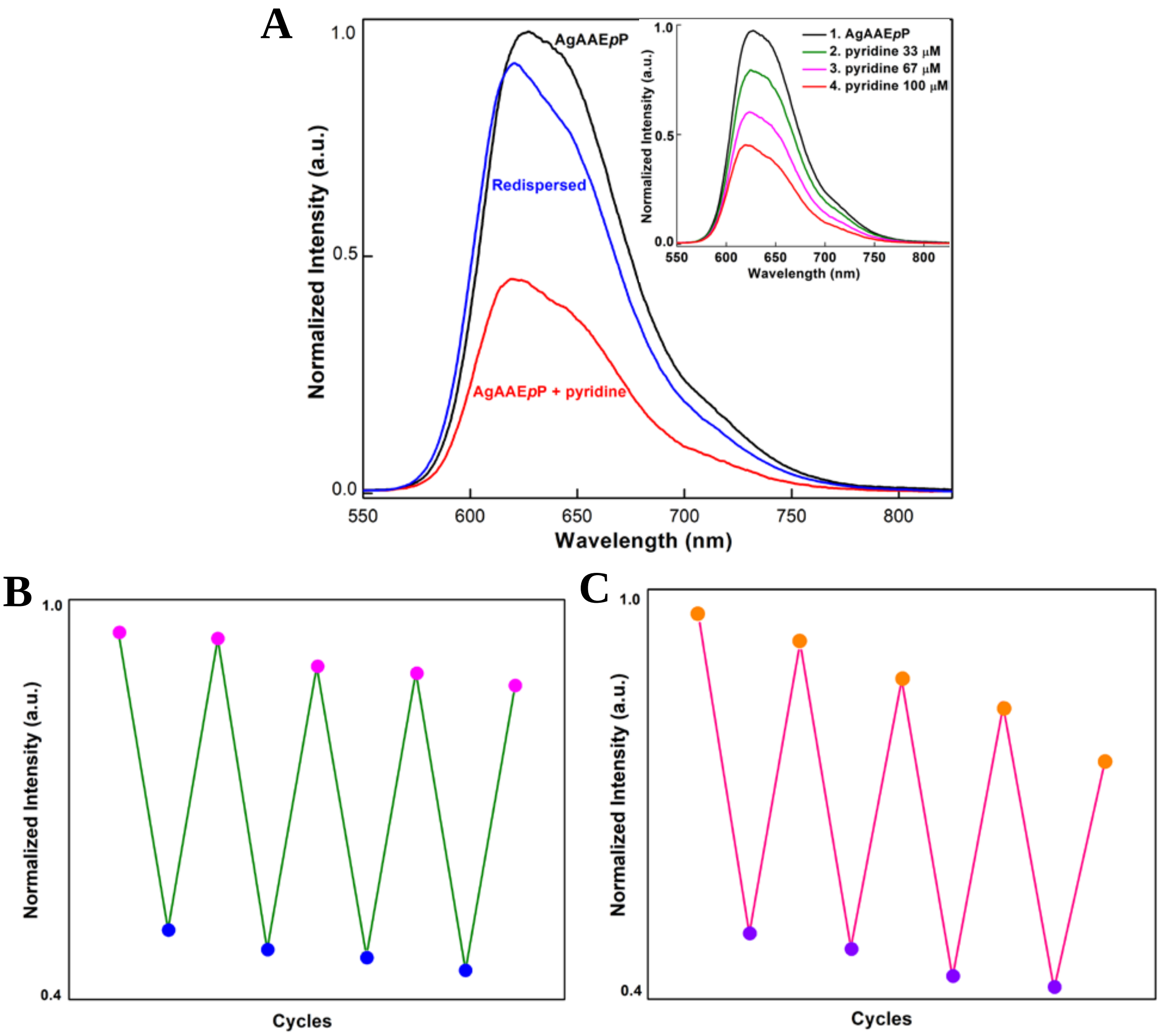

**Figure 8A.** Normalized emission response of AgAAE*p*P to pyridine. AgAAE*p*P (100 μM) was suspended in toluene (black), treated with 100 μM pyridine (red), and redispersed in toluene after removing the solvent and analyte by sparging with $N_2$ (blue). The response also shows stepwise decreases when incremental amounts of pyridine were added to reach final concentrations of 33 μM, 67 μM, and 100 μM. (inset). **8B.** Normalized emission response of AgAAE*p*P to multiple cycles of pyridine addition and removal. AgAAE*p*P (100 μM) was exposed to 100 μM of pyridine followed by a sparging and redispersal process a total of 4 times. **8C.** Normalized emission

response of AgAAE*p*P to multiple cycles of pyridine addition and removal. AgAAE*p*P (100 μM) was exposed to 10 μM of DMA followed by an addition of nitric acid to reach a final concentration of 20 μM $H^+$ a total of 4 times.

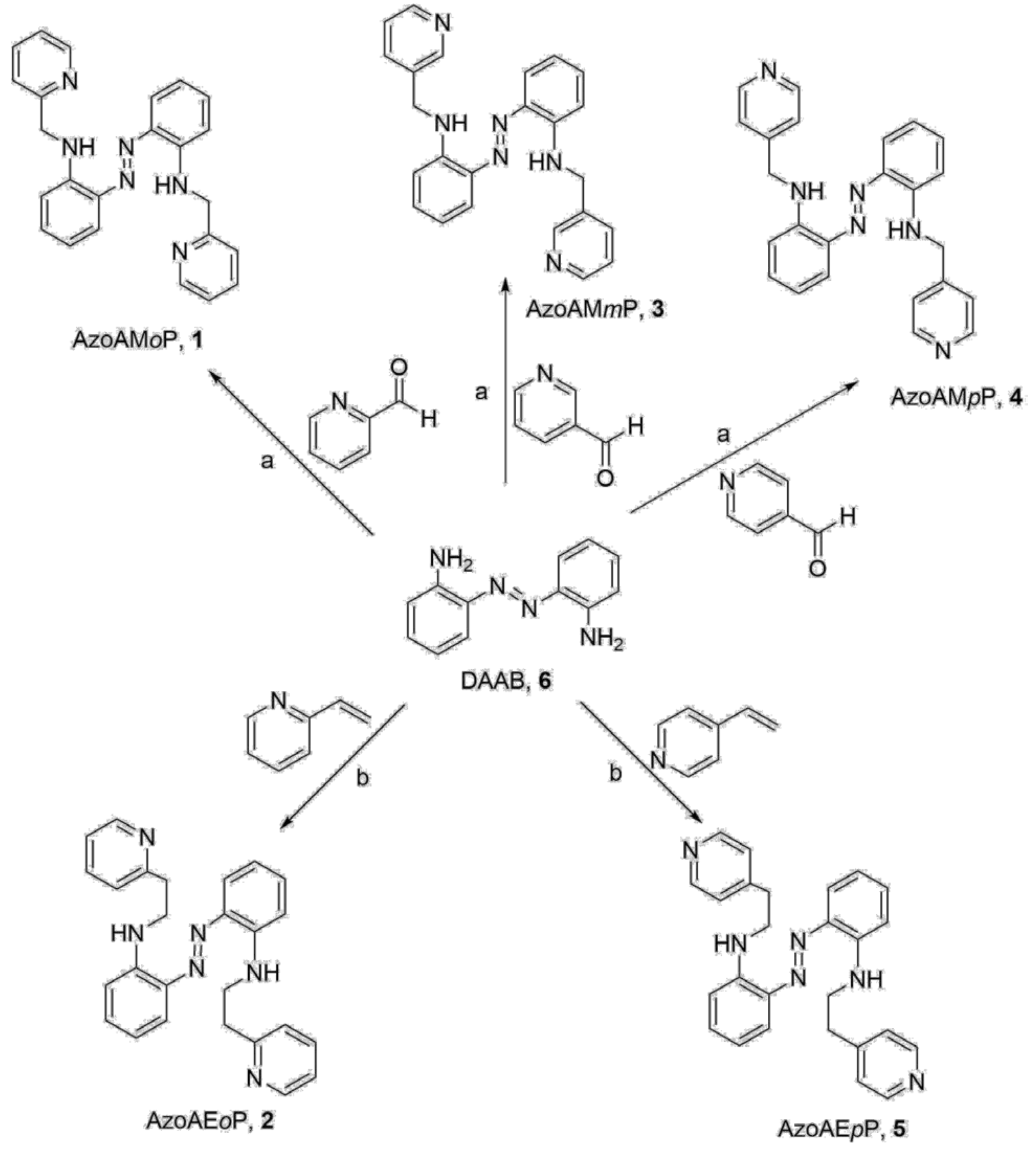

**Scheme 1.** Synthetic protocols for preparing five AzoAX*x*P ligands. Reagents and conditions: (a) $NaBH(OAc)_3$,$CH_2Cl_2$, room temperature; (b) AcOH, MeOH, 45 °C.

**Table 1.** Selected Interatomic Distances (A˚) and Angles (deg) for the five AzoAX*x*P Ligands.

| | Selected bond lengths | Bond angles |
|---|---|---|
| AzoAM*o*P | N(3)–N(3) 1.2793(15)<br>N(3)–C(12) 1.4127(16)<br>C(12)–N(2) 1.3573(18)<br>N(2)–C(6) 1.4434(18) | N(3)–N(3)–C(12) 116.87(10)<br>N(3)–C(12)–C(7) 126.89(11)<br>N(2)–C(7)–C(12) 121.64(11)<br>C(6)–N(2)–C(7) 123.93(11)<br>N(2)–C(6)–C(5) 110.70(10)<br>N(1)–C(5)–C(6) 118.15(12) |
| AzoAM*m*P | N2–N2 1.268(3)<br>C2–N2 1.414(3)<br>C3–N1 1.373(3)<br>C8–N1 1.444(3) | N2–N2–C2 114.9(2)<br>N2–C2–C3 115.07(17)<br>N1–C3–C2 119.75(18)<br>C3–N1–C8 122.48(17)<br>N1–C8–C4 113.39(17) |
| AzoAM*p*P | N1–N1 1.265(3)<br>C7–N1 1.415(2)<br>C12–N2 1.373(2)<br>C1–N2 1.449(2) | N1–N1–C7 115.08(18)<br>N1–C7–C12 115.58(15)<br>N2–C12–C7 119.91(15)<br>C12–N2–C1 122.46(15)<br>N2–C1–C2 115.93(16) |
| AzoAE*o*P | N1–N1 1.267(3)<br>C1–N1 1.408(2)<br>C6–N2 1.354(2)<br>C7–N2 1.446(2)<br>C9–N3 1.340(2) | N1–N1–C1 115.48(18)<br>N1–C1–C6 115.84(15)<br>N2–C6–C1 120.30(16)<br>C6–N2–C7 124.91(17)<br>N2–C7–C8 112.68(15)<br>N3–C9–C8 117.21(17) |
| AzoAE*p*P | N2–N2 1.283(4)<br>N2–C7 1.408(3)<br>N1–C6 1.358(3)<br>N1–C5 1.450(3) | N2–N2–C7 117.2(3)<br>N2–C7–C6 127.1(2)<br>N1–C6–C7 121.0(2)<br>C6–N1–C5 125.0(2)<br>N1–C5–C4 109.4(2) |

**Table 2.** Photophysical Properties of AzoAX*x*P Ligands and Coordination Polymers.

| Compound | H-bond distances NH···N=N(Å) | ε [$M^{-1}cm^{-1}$] | $\lambda_{max}$ (nm) Abs | $Em_{max}$ (nm) | Relative emission intensity (RT) |
|---|---|---|---|---|---|
| AzoAM*o*P | 2.046(15) | 12334 | 495 | 602[a] | n.a. |
| AzoAM*m*P | 2.33(3) | 13729 | 497 | 607[a] | n.a. |
| AzoAM*p*P | 2.34(2) | 13702 | 495 | 603[a] | n.a. |
| AzoAE*o*P | 2.36(2) | 15874 | 501 | 619[a] | n.a. |
| AzoAE*p*P | 2.00(3) | 15195 | 502 | 616[a] | n.a. |
| AgAAM*o*P | 2.43(3), 2.47(5) | n.a. | 440 | 600[b] | 0.035 |
| AgAAM*m*P | 2.12(9), 1.96(2) | n.a. | 440 | 618[b] | 0.067 |
| AgAAM*p*P | 2.02(3) | n.a. | 475 | 590[b] | 0.069 |
| AgAAE*p*P | 2.00(3), 2.05(1), 2.01(3), 2.02(2) | n.a. | 508 | 641[b] | 1.0 |

[a] $\lambda_{ex}$ = 490 nm, [b] $\lambda_{ex}$ = 523 nm

## Supporting Information

The Supporting Information is available free of charge on the ACS Publications website at DOI: XXXXXXXX.

- $^{1}H$ and $^{13}C$ NMR spectra for new AzoAXxP compounds synthesized. TGA and PXRD spectra for all new $Ag^{+}$ coordination polymers and complexes reported.

- Room temperature and 77K emission spectra of all AzoAXxP compounds. Diffuse reflectance and emission spectra of all $Ag^{+}$ coordination polymers. Emission data for analyte additions to AgAAE*p*P.

- Diagrams of $Ag^{+}$–AzoAXxP units showing distance calculations based on the trajectory and positions of the pyridine units.

- Additional computation details, figures and tables of calculated energies and bond lengths

- Complete X-ray tables and fully labeled ORTEP diagrams of all coordination polymers and complexes.

## Accession Codes

CCDC 1854329–1854336 contain the supplementary crystallographic data for this paper. These data can be obtained free of charge via www.ccdc.cam.ac.uk/data_request/cif, or by emailing data_request@ccdc.cam.ac.uk, or by contacting The Cambridge Crystallographic Data Centre, 12 Union Road, Cambridge CB2 1EZ, UK; fax: +44 1223 336033.

AUTHOR INFORMATION

**Corresponding Author**

*scburdette@wpi.edu

**Present Addresses**

[§]Department of Chemistry, University College London, 20 Gordon Street, London, WC1H 0AJ, U.K

**Author Contributions**

The manuscript was written through contributions of all authors. All authors have given approval to the final version of the manuscript.

**ACKNOWLEDGMENTS**

We thank Prof. Christopher Lambert for guidance with diffuse reflectance and solid-state emission measurements. This work was supported by the American Chemical Society Petroleum Research Fund grant 53977-ND3 and Worcester Polytechnic Institute.

SYNOPSIS

Four luminescent silver coordination polymers have been assembled using unique bipyridine ligands containing an azobenzene chromophore. The most emissive complex, AgAAE*p*P, containing tetra-coordinate silver sites, and undergoes reversible disassembly with other silver binding ligands to create a luminescent switch.